\documentclass[pra,showpacs,twocolumn]{revtex4}
\usepackage{bm,graphicx,amsmath}

\begin{document}

\title{Loading of bosons in optical lattices into the $p$ band}
\author{Jonas Larson}
\email{jolarson@fysik.su.se}
\affiliation{Department of Physics, Stockholm University, Se-106 91, Stockholm, Sweden}
\author{Jani-Petri Martikainen}
\affiliation{NORDITA, Se-106 91
Stockholm, Sweden}

\date{\today}

\begin{abstract} 
We present a method for transferring bosonic atoms residing on the lowest $s$-band of an optical lattice to the first excited $p$-bands. Our idea hinges on resonant tunneling between adjacent sites of accelerated lattices. The acceleration effectively shifts the quasi-bound energies on each site such that the system can be cast into a Wannier-Stark ladder problem. By adjusting the acceleration constant, a situation of resonant tunneling between the $s$- and $p$-bands is achievable. Within a mean-field model, considering $^{87}$Rb atoms, we demonstrate population transfer from the $s$- to the $p$-bands with around 95 $\%$ efficiency. Nonlinear effects deriving from atom-atom interactions, as well as coupling of the quasi bound Wannier-Stark states to the continuum, are considered.  
\end{abstract}

 \maketitle

\section{Introduction} 
With the recent experimental progress, systems of ultracold atoms in optical lattices have become a principle playground for the study of correlated many-body systems typically appearing in condensed matter theories~\cite{maciek}. For example, the ability to cool down atomic gases to very low temperatures, having access to the potential parameters as well as to the atom-atom interaction strength, made it possible to detect, for the first time experimentally, the Mott-superfluid phase transition in these systems~\cite{bloch}. 

To date, most of the experiments have been confined to the lowest energy band of the optical lattice, the so-called $s$-band, and effects arising due to higher bands could be safely ignored. Contrary to the $s$-band, the onsite wave-functions, or Wannier functions, on the first excited bands, the $p$-bands, have a single node in some spatial direction giving them an orbital character. In the direction of the node, the onsite functions are more extended and therefore the tunneling coefficients in this direction is greater than in the perpendicular directions. Due to this anisotropic structure, the physics on excited bands allow for new phenomena not present on the lowest band, like supersolids or other types of novel phases~\cite{pband}. Moreover, the $p$-bands are two-fold or three-fold degenerate (in two or three dimensions respectively), and consequently enable non-trivial phases even within the insulating phases~\cite{jonas3}. 

In fermionic systems, the $p$-bands can be excited simply by considering filling factors larger than unity such that the lower $s$-band is completely filled~\cite{jani1}. For ultracold bosons, which we consider in this work, populating the $p$-bands implies an excited non-equilibrium state and the system is thereby energetically not stable. Even so, it turns out that the life-time for bosons on the $p$-bands is long in comparison to characteristic time-scales for tunneling between neighboring lattice sites~\cite{jonas3,jani2}. Thus, despite decay, non-trivial physics can indeed be experimentally realized on the $p$-bands, for example condensation into non-zero momentum states has already been seen~\cite{phillips,hemmerich1,hemmerich2}. The experimental methods used for preparing bosons on the $p$-bands include; Raman transitions via a third band~\cite{raman}, adiabatically ramping up a constantly moving lattice~\cite{phillips}, or sudden switching of superlattices~\cite{hemmerich1,hemmerich2}. In the last scheme, loading of both $p$-band bosons~\cite{hemmerich1} and $f$-band bosons~\cite{hemmerich2} were achieved.

In the present work we analyze, on a mean-field level, an idea somewhat similar to the one employed in Refs.~\cite{hemmerich1,hemmerich2}, but for a regular monochromatic lattice, i.e. a lattice with only one characteristic wave-length $\lambda$. We suggest a sudden tilting of the lattice 
via acceleration of it~\cite{phillips,BO,boost}. 
The dynamics in the tilted system is conveniently studied in the Wannier-Stark picture where we form a basis of localized states at each lattice site. The corresponding energies of these states are the Wannier-Stark energies, which form resonances in the full continuous spectrum of the system. In the regime of deep lattices, we impose a tight-binding approximation where transitions appear only between neighboring lattice sites. By adjusting the tilting, a resonance coupling between the $s$- and the $p$-bands occurs. Suddenly switching off the acceleration/tilting after half a Rabi period leaves most of the initial $s$-band atoms in the $p$-bands. Transitions into higher excited bands are suppressed due to the off-resonant transitions with these bands. Including the $s$-, $p$-, and $d$-bands, we consider the time-dependent Hamiltonian which takes the switch on/off into account. We analyze both the one and the two dimensional cases.
It is furthermore shown that interaction between the atoms leads to a self-trapping effect and hence a decrease in the loading efficiency. This, however, becomes important only for very strongly interacting systems and should therefore not be a problem as long as one considers a weakly interacting gas. Throughout we use experimentally relevant parameters and demonstrate a transfer of atoms from the $s$- to the $p$-bands of more than 95 $\%$.

\section{Model System}
\subsection{Multi-band Wannier-Stark ladders}
The spectrum of a Schr\"odinger operator on the form
\begin{equation}\label{ham1}
H_0=-\frac{\hbar^2}{2m}\frac{d^2}{dx^2}+V_L\sin^2(kx)
\end{equation}
has a band-structure $E_\nu(q)$ characterized by a discrete band index $\nu=1,\,2,\,3,\,...$ and a quasi momentum $-2\pi/\lambda\leq q<2\pi/\lambda$. Here, $V_L$ is the potential depth, $m$ will be the mass or our atoms, and $k=2\pi/\lambda$ the wave number where $\lambda$ is the lattice wave-length. For the lattices that we consider in the present paper, we label the first three bands $s$, $p$, and $d$ respectively~\cite{bands}. For higher dimensions there are multiple $p$- and $d$-bands. Whenever the depth of the lattice potential is increased, the band-widths $|E_\nu(0)-E_\nu(2\pi/\lambda)|$ decrease. In the limit of infinite depth, the bands become completely flat and relative to the onsite oscillator energy approximately equidistant, mimicking the spectrum of a harmonic oscillator $E_\nu\equiv\left(4\pi/\lambda\right)^{-1}\int_{Br}dq\,E_\nu(q)\propto\nu$. In such a situation, particles residing in the lattice are immobile. 

\begin{figure}[ht]
\begin{center}
\includegraphics[width=8cm]{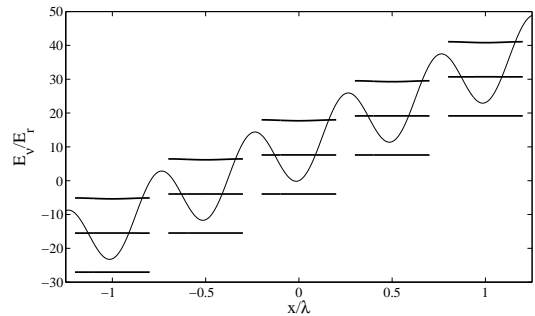}
\caption{A tilted periodic sine potential with its corresponding Wannier-Stark energies (here we have not averaged the energies over the Brillouin zone in order to demonstrate the band widths). The potential depth is $V_L=20E_r$. } \label{fig1}
\end{center}
\end{figure}

For a constant force with strength $F_0$ applied to our system,
\begin{equation}
H_1=-\frac{\hbar^2}{2m}\frac{d^2}{dx^2}+V_L\sin^2(kx)-F_0x,
\end{equation} 
it is practical to define the Wannier-Stark energies~\cite{ws} which approximate
\begin{equation}
\varepsilon_{\nu,j}=E_\nu+jF_0\frac{\lambda}{2}-i\Gamma_\nu,
\end{equation}
with $j$ being an integer labeling the sites and $\Gamma_\nu$ gives the decay of the $\nu$'th band into the continuum of states. For each $\nu$, the energies form a so called Wannier-Stark ladder. Given $j$, the low lying energies correspond to quasi bound states localized at site $j$. A schematic picture of the Wannier-Stark energies is presented in Fig.~\ref{fig1} showing the first three bands/ladders. In this example, we have taken $V_L=20E_r$, where $E_r=\hbar^2k^2/2m$ is the recoil energy. The linear force is taken such that for the real parts $\varepsilon_{2,j}=\varepsilon_{1,j+1}$, i.e 
\begin{equation}\label{forceamp}
F_0=\frac{2(E_2-E_1)}{\lambda}c,
\end{equation}
where $c$ is a dimensionless parameter that we have introduced for later purposes. More precisely, in the full system an infinite number of states are coupled while only a fraction of them couple resonantly. Population transfer occurs predominantly among these resonant states and off-resonant states can be, in a first approximation, ignored. Nevertheless, the virtual processes related to coupling to far detuned states induce small energy shifts in the states of interest, and consequently shift their mutual resonance conditions. An expression for these shifts can, in principle, be obtained by adiabatic elimination of off-resonant states, but this is not manageable for the system sizes we study, and instead we introduce the constant $c$ in (\ref{forceamp}). In other words, when necessary $c$ adjusts for the virtual energy shifts originating from off-resonant processes. In the figure $c=1$. For these parameters, the $s$- and $p$-bands are almost completely flat while the $d$-band has a small width. Even though the bands (especially the $d$-band) have a small width, neighboring $p$- and $d$-bands are mutually off resonant for any quasi momentum $q$.

The imaginary part $\Gamma_\nu$ of the Wannier-Stark energies can be made relatively small for the $s$- and $p$-bands assuming sufficiently deep lattices, as for example in Fig.~\ref{fig1}. When the coherent coupling between the two resonant states are large compared to the decay rate, the two states can perform numerous of Rabi oscillations before they decay. This is the present idea for loading of bosons into the first excited bands. The energies $E_\nu$ can be obtained by direct diagonalization of the untilted Hamitonian. The imaginary parts $\Gamma_\nu$ can either be found by diagonalization of the full evolution operator~\cite{ws} or estimated using Landau-Zenner theory~\cite{ws,lz}. For our set of parameters, we found it numerically difficult to get $\Gamma_\nu$ utilizing these methods. Instead we estimate $\Gamma_\nu$ for the two lowest $s$- and $p$-bands by performing a wave-packet propagation of the full Hamiltonian and calculate the decay of population out from the first two Brillouin zones~\cite{jonasbo}.

\subsection{Tight-binding model for accelerated lattices}
As a basis for our analysis, we will consider $^{87}$Rb atoms with mass $m$ ($=1.44\times10^{-25}$ kg) trapped in an isotropic harmonic trap with frequency $\omega=40\times2\pi$ Hz. The atoms are exposed to an optical lattice, with the wavelengths $\lambda=795$ nm and the depths $V_L=20E_r$ the same in all spatial directions~\cite{com2}. These are typical experimental parameters, even though the lattice depth is relatively deep, however still well within experimental reach. We have found numerically that a deep lattice is preferable for our loading scheme.  Only the one- and two dimensional cases will be considered in this work (the generalization to three dimensions is straightforward, but numerically demanding). The derivation of the Hamiltonian will be for the two dimensional case. The lattice is formed by counter propagating waves in such a way that the relative frequency $\Delta\omega_L(t)$ of the two waves can be tuned in time $t$. Such lattices, with $\Delta\omega(t)=k\left(at^2-bt\right)$ ($a$ and $b$ are constants), are typically used for the study of Bloch oscillations of cold atoms~\cite{BO}. Within the mean-field regime, and assuming $s$-wave scattering between the atoms, the equation of motion for the order parameter, after a proper unitary transformation~\cite{boost}, reads
\begin{equation}\label{eom1}
\begin{array}{lll}
\displaystyle{i\hbar\frac{\partial}{\partial t}\psi(\mathbf{x},t)} & = &
\displaystyle{ \left[-\frac{\hbar^2}{2m}\nabla^2+V_{tr}(\mathbf{x})+V_{lat}(\mathbf{x})+F\mathbf{x}\right.}\\ \\
& & \displaystyle{\left.+\frac{U_0}{2}|\psi(\mathbf{x},t)|^2\right]\psi(\mathbf{x},t)},
\end{array}
\end{equation}   
where $F=ma$ is the amount of tilting of the lattice, and $V_{lat}(\mathbf{x})=V_L\left[\cos^2(kx)+\cos^2(ky)\right]$. 

Initially we will assume $F=0$ and the atoms to be prepared in the lowest $s$-band. The acceleration of the lattice is switched on during a time $T_s$ and stays on for a time $T_r$ and then switched off during $T_s$ again. To mimic such an experimental ramping, for our numerics we take the force as
\begin{equation}\label{pulse}
F(t)=\frac{F_0}{2}\left[\tanh\left(\frac{t-0.1T_r}{T_s}\right)-\tanh\left(\frac{t-1.1T_r}{T_s}\right)\right]
\end{equation}
with $F_0$ given in Eq.~(\ref{forceamp}). This $F(t)$ has a smoothened square pulse shape, starting at $t=0.1T_r$ and ending at $t=1.1T_r$, and with a turn on/off time $T_s$. As explained above, the parameter $F_0$ and the time $T_r$ are chosen such that the lowest Wannier-Stark energy is resonant with the first excited states of a neighboring site, and the system undergoes half a Rabi cycle. Due to the anharmonicity of the lattice potential, the second excited Wannier-Stark energies will be off-resonant with neighboring first excited Wannier-Stark energies. Nevertheless, for small detunings, higher bands may still be populated and in our calculations we therefore include all the $s$-, $p$-, and $d$-bands. In two dimensions, the first excited bands are two-fold degenerate and the second excited bands three-fold degenerate. We will specifically pick the lattice depth to be 20 recoil energies, $V_L=20E_r$. In this case the low lying bands are relatively flat, but still the anaharmonicity is present. This was demonstrated in Fig.~\ref{fig1}.

Apart from restricting the numerics to contain only the $s$-, $p$-, and $d$-bands, we will further impose the tight-binding approximation. This means that we consider hopping only between neighboring sites. Both these approximations should be justified for relatively large lattice depths $V_L$. As the localized Wannier-Stark states at each lattice site $\mathbf{i}=(i_x,i_y)$ we take the numerically obtained Wannier states~\cite{am}. This is a good approximation for the low lying states of deep lattices~\cite{ws}. In order to derive mean-field equations of motion in the restricted Hilbert space, we expand the condensate order parameter as
\begin{equation}
\psi(\mathbf{x},t)=\sum_\alpha\sum_\mathbf{i}\varphi_\mathbf{i}^\alpha(t)w_\mathbf{i}^\alpha(\mathbf{x}),
\end{equation}
where $\alpha=0,\,x,\,y,\,xy,\,x0,\,0y$ and $\mathbf{i}$ runs over all lattice sites. The $\alpha$-index represents six possible Wannier states ($s$-band: ``0'', $p$-band: ``x'' and ``y'', and $d$-band: ``xy'', ``0x'', and ``0y''). Normalization is such that $N_0=\int\,d\mathbf{x}\,|\psi(\mathbf{x},t)|^2$, where $N_0$ is the total number of atoms. The energy expressed in the coefficients $\varphi_\mathbf{i}^\alpha(t)$ becomes, 
\begin{equation}\label{energy}
\begin{array}{lll}
E & = & \displaystyle{\sum_\alpha\sum_\mathbf{i}\left(\varepsilon_{\alpha,\mathbf{i}}+V_{tr}(\mathbf{i})\right)|\varphi_\mathbf{i}^\alpha|^2}\\ \\
& & \displaystyle{+\sum_{\alpha,\beta}\sum_{\langle\mathbf{i},\mathbf{j}\rangle}t_{\mathbf{i}\mathbf{j}}^{\alpha\beta}\left(\varphi_\mathbf{i}^{*\alpha}\varphi_\mathbf{j}^\beta+\varphi_\mathbf{j}^{*\beta}\varphi_\mathbf{i}^\alpha\right)}\\ \\ & & \displaystyle{+\frac{1}{2}\sum_{\alpha,\beta,\gamma,\delta}\sum_\mathbf{i}U_{\alpha\beta\gamma\delta}\varphi_\mathbf{i}^{*\alpha}\varphi_\mathbf{i}^{*\beta}\varphi_\mathbf{i}^\gamma\varphi_\mathbf{i}^\delta},
\end{array}
\end{equation}
where
\begin{equation}\label{overlap}
\begin{array}{c}
\displaystyle{t_{\mathbf{i}\mathbf{j}}^{\alpha\beta}=\!\int d\mathbf{x}\,w_\mathbf{i}^\alpha(\mathbf{x})\!\left[-\frac{\hbar^2}{2m}\nabla^2\!+V_{lat}(\mathbf{x})+F\mathbf{x}\right]\!w_\mathbf{j}^\beta(\mathbf{x})},\\ \\
\displaystyle{U_{\alpha\beta\gamma\delta}=U_{scat}\int\,d\mathbf{x}\,w_\mathbf{i}^\alpha(\mathbf{x})w_\mathbf{i}^\beta(\mathbf{x})w_\mathbf{i}^\gamma(\mathbf{x})w_\mathbf{i}^\delta(\mathbf{x})}.
\end{array}
\end{equation}
The summation $\langle\mathbf{i},\mathbf{j}\rangle$ runs over nearest neighbors, we have assumed the atom-atom interaction to be non-zero only within one and the same site, and asterix $*$ denotes complex conjugation. The adjustable parameter $U_{scat}$ is determined by the system dimensions and the $s$-wave scattering length $a_s$. The effect of the trapping potential is seen only in the shift of the onsite energies. This should be justified since the slope of $V_{tr}(\mathbf{x})$ is typically a thousand times smaller than $|F_0|$ within the atomic cloud and should therefor not induce any hopping between sites. We further note that the lattice Hamiltonian $H_{lat}=\mathbf{p}^2/2m+V_{lat}(\mathbf{x})$ only couples Wannier functions within the same band, i.e. $\langle w_\mathbf{i}^\alpha|H_{lat}|w_\mathbf{j}^\beta\rangle=0$ whenever $\alpha\neq\beta$, and as a consequence it is the tilting alone that induces the transition from the $s$- to the $p$-bands.

\section{Loading in a one dimensional system}
The expression for the energy (\ref{energy}) is general for any dimensions. In one dimension, the excited bands are not degenerate and we can label them by $\alpha=s,\,p,\,d$. As we will demonstrate, the underlying physics is easily understood in one dimension. Nevertheless, some of the effects are more significant in two compared to one dimensions and vice versa.

\subsection{Two-site problem and the effect of interactions}

\begin{figure}[ht]
\begin{center}
\includegraphics[width=8cm]{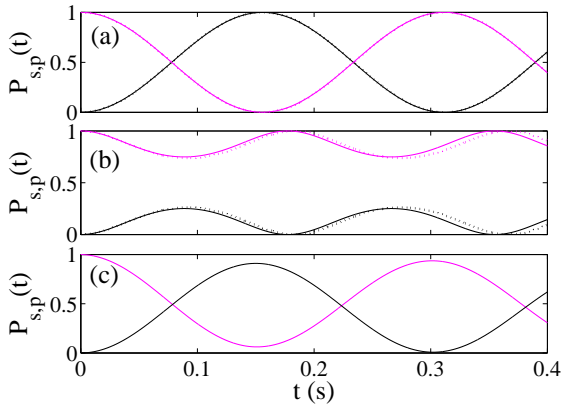}
\caption{Populations $P_{s,p}(t)$ in the $s$- (pink curves) and $p$-band (black curves). In the upper two plots we compare the results obtained from the two-state effective model, Eq.~(\ref{jj2}) (dotted curves), with those obtained from the two-site six-state model defined in Eq.~(\ref{2s}) (solid curves). In (a) $U_s=U_p=0$ and in (b) $U_sN_0\approx4.3t_{12}^{sp}$ and $U_pN_0\approx3.1t_{12}^{sp}$. The effect of the interaction is manifested as self-trapping. The lower plot (c) displays the full $N$-site model ($N=41$) including a trapping potential. The interaction is set to zero in this case. In all examples $V_L=20E_r$, the tilt parameter $c=1$, and for the trap frequency in (c) $\omega=40\times2\pi$ Hz. } \label{fig2}
\end{center}
\end{figure}

Effects deriving from atom-atom interactions, as well as the dynamics in general, can be extracted by considering the most simple situation consisting of a lattices with only two sites similar to a double-well system. Here we consider the static model where the force $F(t)=F_0$ is time-independent such that we expect Rabi evolution. Thus, we have a set of six coupled equations for the coefficients $\varphi_1^s(t),\,\varphi_1^p(t),\,\varphi_1^d(t),\,\varphi_2^s(t),\,\varphi_2^p(t),\,\varphi_2^d(t)$. The atom-atom interaction-free Hamiltonian $H_2$ reads
\begin{equation}\label{2s}
H_2=\left[\begin{array}{cccccc}
0 & 0 & 0 & t_{12}^{ss} & t_{12}^{sp} & t_{12}^{sd}\\
0 & \Delta_{21} & 0 & t_{12}^{sp} & t_{12}^{pp} & t_{12}^{pd}\\
0 & 0 & \Delta_{31} & t_{12}^{sd} & t_{12}^{pd} & t_{12}^{dd}\\
t_{12}^{ss} & t_{12}^{sp} & t_{12}^{sd} & -\Delta_{21} & 0 & 0\\
t_{12}^{sp} & t_{12}^{sp} & t_{12}^{pd} & 0 & 0 & 0\\
t_{12}^{sd} & t_{12}^{pd} & t_{12}^{dd} & 0 & 0 & \Delta_{32}
\end{array}\right],
\end{equation}
where we have introduced the detunings $\Delta_{ij}=\varepsilon_{1,i}-\varepsilon_{2,j}$. Only the $\varphi_1^s(t)$ and the $\varphi_2^p(t)$ states are degenerate. Moreover, $|t_{12}^{\alpha\beta}|$ is typically 2-4 orders of magnitude smaller than any of the detunings for relevant lattice depths ($V_L\sim20E_r$). Thereby, all couplings apart from the one being resonant is well in the dispersive regime and should only contribute with virtual processes. In other words, it is justified to introduce an effective $2\times2$ model for the $\varphi_1^s(t)$ and $\varphi_2^p(t)$ states alone;
\begin{equation}\label{jj}
i\hbar\frac{\partial}{\partial t}\left[\begin{array}{c}
\varphi_1^s(t)\\
\varphi_2^p(t)\end{array}\right]=\left[\begin{array}{cc}
0 & \tilde{t}_{12}^{sp}\\
\tilde{t}_{12}^{sp} & 0\end{array}\right]
\left[\begin{array}{c}
\varphi_1^s(t)\\
\varphi_2^p(t)\end{array}\right],
\end{equation}
showing perfect Rabi oscillations between the two states. The inverse Rabi frequency $\tilde{t}_{12}^{sp}$ is the effective coupling obtained when all the dispersive states have been adiabatically eliminated, i.e. in lowest order (in terms of coupling strengths divided by detunings) we have $\tilde{t}_{12}^{sp}=t_{12}^{sp}$. 

Equation (\ref{jj}) is the well-known Rabi model and is often considered for bosonic Josephson systems. It is known that interaction among the bosons (atoms) may greatly affect the Josephson dynamics in terms of self-trapping~\cite{smerzi}. Assuming lowest order in the coupling parameter and including onsite atom-atom interaction results in the effective model
\begin{equation}\label{jj2}
i\hbar\frac{\partial}{\partial t}\left[\begin{array}{c}
\varphi_1^s(t)\\ \\
\varphi_2^p(t)\end{array}\right]=\left[\begin{array}{cc}
U_s|\varphi_{1}^s(t)|^2 & \tilde{t}_{12}^{sp}\\ \\
\tilde{t}_{12}^{sp} & U_p|\varphi_2^p(t)|^2\end{array}\right]
\left[\begin{array}{c}
\varphi_1^s(t)\\ \\
\varphi_2^p(t)\end{array}\right],
\end{equation}
where $U_s=U_{ssss}$ and $U_p=U_{pppp}$ are the onsite effective scattering coefficients. Whenever the diagonal elements $U_s|\varphi_1^s(t)|^2$ and $U_p|\varphi_2^p(t)|^2$ are different, the dynamics is efficiently detuned and the population transfer is not perfect. Thus, the interaction can been seen as inducing effective shifts of the onsite energies. This self-trapping effect is demonstrated in Fig.~\ref{fig2} displaying the populations $P_\alpha(t)=|\varphi_\alpha(t)|^2$; in (a) the atom-atom interaction is zero while in (b) it is non-zero. We show the numerical solutions of the two-site six-state model and the solutions of the effective two-state model (\ref{jj2}). Without interaction, the agreement between the two models is almost perfect and therefore the transfer is approximately 100 $\%$. The results disagree somewhat when the interaction is non-zero, even if the qualitative structure is the same. Within the six-state model we included all non-zero interaction terms $U_{\alpha\beta\gamma\delta}$. The effective interaction strength $U_sN_0$ used in (b) is very large in order to demonstrate the self-trapping phenomena. In fact, we have in this example $U_sN_0\approx4.8|t_{12}^{ss}|$ implying the strongly interacting regime. For example, the critical coupling for entering the Mott insulator state with one atom per site $U_sN_0\approx4|t_{12}^{ss}|$~\cite{sc}. For a factor 10 smaller coupling, the six-state model predicts a transfer of 96.7 $\%$, and for a factor 100 the transfer is 98.8 $\%$. For typical experimental parameters it is hence expected that the self-trapping effect will be vanishingly small, and in the following subsection we will focus on a non-interacting gas.

\subsection{Full one dimensional dynamics}

\begin{figure}[ht]
\begin{center}
\includegraphics[width=8cm]{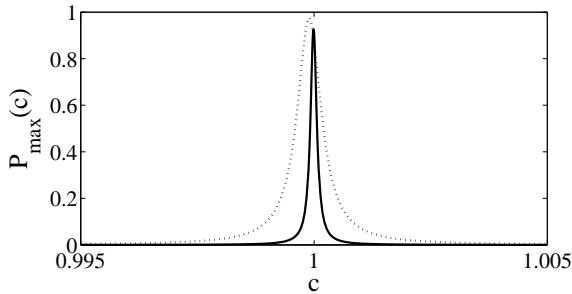}
\caption{Maximum population transfer $P_{max}(c)$ as a function of the dimensionless parameter $c$. The solid line is the result for one dimension and the dotted for two dimensions. The potential depth $V_L=20E_r$ and $\omega=40\times2\pi$ Hz.} \label{fig3}
\end{center}
\end{figure}

We showed above that in the two-site system, an effective two-level model qualitatively describes the dynamics. In a $N$-site system there are numerous of participating levels, even though the full Hamiltonian is six-diagonal. Adiabatically eliminating all far detuned levels might therefore induce non-vanishing shifts in the resulting effective two-level model. Assuming a two-level model to be valid, we modify Eq.~(\ref{jj}) to allow for a detuning between the two levels
\begin{equation}\label{jj3}
\begin{array}{c}
i\hbar\frac{\partial}{\partial t}\left[\begin{array}{c}
\varphi_1^s(c,t)\\ \\
\varphi_2^p(c,t)\end{array}\right]=\\ \\
\left[\begin{array}{cc}
\frac{E_2-E_1}{2}(\delta_0-c) & g\\ \\
g & -\frac{E_2-E_1}{2}(\delta_0-c)\end{array}\right]
\left[\begin{array}{c}
\varphi_1^s(c,t)\\ \\
\varphi_2^p(c,t)\end{array}\right].
\end{array}
\end{equation}
Thus, if the full $N$-site Hamiltonian shows similar dynamics as predicted by (\ref{jj3}) we conclude that the system Hamiltonian is effectively $2\times2$ block diagonal. Moreover, for vanishing shifts $g=t_{12}^{sp}$ and $\delta_0=1$ so that $c=1$ result in a vanishing effective detuning (remember that $c$ was introduced in order to adjust for any virtual energy shifts). Denoting $\Delta(c)=(E_2-E_1)(\delta_0-c)/2$, the maximum population transfer $P_{max}(c)$ to the $p$-band, provided $\varphi_s(c,t=0)=1$, is given by the Lorentzian
\begin{equation}
P_{max}(c)=\frac{g^2}{\Delta^2(c)+g^2}
\end{equation}
having a width $\sim g$. In Fig.~\ref{fig3} (solid line), we test this hypothesis by calculating the maximum population transfer for the $N=41$ site model including all $6N$ states and a harmonic trap. For the numerics, we truncate the lattice such that the atomic density is approximately zero (due to the confining trap) at the edge of the lattice. The initial state is taken as the ground state for the $F=0$ case. $P_{max}(c)$ in Fig.~\ref{fig3} shows a clear Lorentzian shape indicating that the dynamics could be effectively described by a two-level model. Corrections are manifested in, for example, the fact that $P_{max}(c)$ does not exactly reach 1. The offset $\delta_0$ ($\approx0.99998$) from unity gives the shift in the onsite energies, while the width determines the effective coupling between the $s$- and the $p$-band. Fitting a Lorentzian to the data of Fig.~\ref{fig3} gives a coupling $g\approx0.99t_{12}^{sp}$. The fact that $g$ is only slightly shifted from $t_{12}^{sp}$ and that $\delta_0$ approximates unity indicate minimal virtual shifts arising from off-resonant levels. 

\begin{figure}[ht]
\begin{center}
\includegraphics[width=8cm]{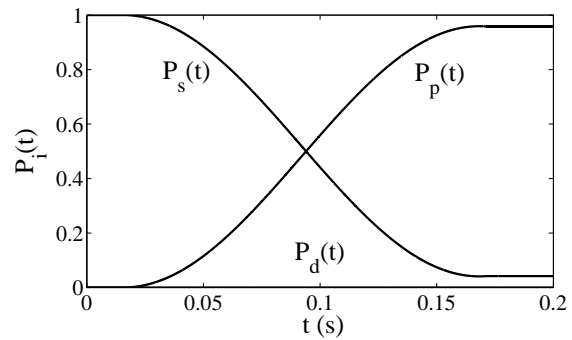}
\caption{Populations $P_i(t)$ ($i=s,\,p,\,d$) for the pulsed acceleration (\ref{pulse}) with $T_s=10$ $\mu$s and $T_r=2\pi\hbar/|t_{ii+1}^{sp}|$. The number of sites was taken as $N=41$, well beyond the extent of the atomic density. The potential depth $V_L=20E_r$, $c=1$, the loss rates $\Gamma_\nu/\hbar=0.1$ Hz ($\nu=s,\,p,\,d$), and $\omega=40\times2\pi$ Hz, while the interactions are set to zero. The population transfer from the $s$- to the $p$-band is $\sim95.5$ $\%$.  } \label{fig4}
\end{center}
\end{figure}

This far, we have not included any coupling of the resonances to the continuum, i.e. taking the imaginary part $\Gamma_\nu$ of the Wannier-Stark energies into account. Utilizing a wave-packet propagation method~\cite{jonasbo}, we estimate $\Gamma_\nu$ by calculating the decay of population within the first two Brillouin zones as time progresses. This gives us a decay rate $\tau=\Gamma/\hbar\approx0.1$ $s^{-1}$, giving a survival time ($\sim10$ s) which is considerably larger than the characteristic Rabi time ($\sim0.1$ s). Another aspect that we have not been investigated is the effects deriving from switching on/off of the lattice acceleration. Figure \ref{fig2} shows that half a Rabi period is approximately given by $T_r\approx2\pi\hbar/|t_{12}^{sp}|$, which we will use for the pulse duration in the expression for $F(t)$ in Eq.~(\ref{pulse}). For minimizing the loading time, $T_s$ should be chosen small. However, a too small switch on/off time $T_s$ may cause non-adiabatic excitations. We pick $T_s=10$ $\mu$s, giving $F(t)$ a square pulse shape with rounded edges. The result of our numerical solution of the full one dimensional system including a trapping potential and a pulse shaped acceleration is presented in Fig.~\ref{fig4}. The final population on the $p$-band is 95 $\%$. The effect of the imaginary coupling to the continum is a drop of the population transfer by approximately 1 $\%$.

\section{Loading in a two dimensional system}

The method used for the one dimensional loading transforms directly to higher dimensions. In Fig.~\ref{fig3}, the dotted curve displays the maximum population transfer to the $p$-bands in two dimensions. The initial state is again the ground state for $F=0$ and the trap frequency $\omega=40\times2\pi$ Hz. The shape is still Lorentzian, and the shift $\delta_0\approx0.99990$. However, the effective coupling is roughly four times as large as in one dimension (a Lorentzian fit gives $g\approx3.57t_{12}$), which implies a speed-up of the $p$-band loading.

One difference between the two cases is that in two dimensions there are two $p$- and three $d$-bands. In addition, atoms can hop in two directions, $x$ and $y$. If an atom in a $s$-state hops to a $p$-state orbital in the $x$-direction, due to symmetry it can only hop into a $p$-state orbital with a node in the $x$-direction. Similarly for hopping in the $y$-direction. Thus, neglecting all off-resonant transitions and atom-atom interaction one derives an effective three-state model
\begin{equation}
i\hbar\frac{\partial}{\partial t}\left[
\begin{array}{c}
\varphi_1^0(t) \\
\varphi_2^x(t)\\
\varphi_3^y(t)
\end{array}\right]=\left[
\begin{array}{ccc}
0 & t_{12}^{0x} & t_{13}^{0y}\\
t_{12}^{0x} & 0 & 0\\
t_{13}^{0y} & 0 & 0
\end{array}\right]\left[
\begin{array}{c}
\varphi_1^0(t) \\
\varphi_2^x(t)\\
\varphi_3^y(t)
\end{array}\right],
\end{equation}
where we have labeled the three sites $1,\,2,\,3$. By defining $\varphi^s(t)=\varphi_1^0(t)$ and $\varphi_\pm^p(t)=(\varphi_2^x(t)\pm\varphi_3^y(t))/\sqrt{2}$ and using $t_{12}^{0x}=t_{13}^{0y}\equiv t^{sp}$ we find an effective two-site model
\begin{equation}\label{jj4}
i\hbar\frac{\partial}{\partial t}\left[\begin{array}{c}
\varphi^s(t)\\
\varphi_+^p(t)\end{array}\right]=\left[\begin{array}{cc}
0 & \sqrt{2}t^{sp}\\
\sqrt{2}t^{sp} & 0\end{array}\right]
\left[\begin{array}{c}
\varphi^s(t)\\
\varphi_+^p(t)\end{array}\right].
\end{equation}
Thus, comparable to the one dimensional loading, here the coupling has been increased by a factor $\sqrt{2}$. However, Fig.~\ref{fig3} made clear that the effective coupling is in fact increased by a factor $\sim4$ rather than $\sqrt{2}$. This is assumed to derive from the fact that there are many more off-resonant levels and the corresponding virtual processes influence the effective coupling more drastically when all of them are eliminated. 
 
Since there are multiple $p$- and $d$-bands we define the total populations as
\begin{equation}
\begin{array}{ll}
P_s(t)=P_0(t), & \mathrm{{\it s}-band}\\ \\
P_p(t)=P_x(t)+P_y(t), & \mathrm{{\it p}-band}\\ \\
P_d(t)=P_{xy}(t)+P_{x0}(t)+P_{0y}(t), & \mathrm{{\it d}-band}.
\end{array}
\end{equation}
We will mainly be interested in the transfer efficiency and not in the actual quantum state loaded in the p-bands. This can be motivated from the fact that the prepared state is not a stationary one~\cite{jonas3}, and non-trivial dynamics continues also when the acceleration has been turned off. For example, in the experiment of Ref.~\cite{hemmerich1} it was shown how condensation appeared on the $p$-bands after loading into these bands.

\begin{figure}[ht] 
\begin{center}
\includegraphics[width=8cm]{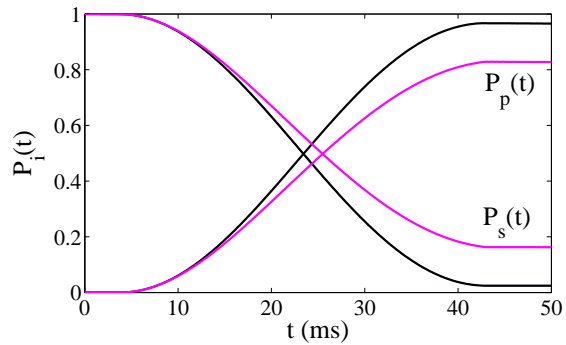}
\caption{Probabilities $P_s(t)$ and $P_p(t)$. The final transfer to the $p$-bands is 96.5 $\%$ in the non-interacting case (black curves) and 83.0 $\%$ for the interacting situation (pink curves). The parameters are the same as in Fig.~\ref{fig4}, but the dimensionless parameter $c=0.9999$, and the half Rabi oscillation time was picked as $T_r=39$ ms. The interaction strengths have been taken very high in order to achieve a self-trapping effect; $U_{0000}N_0\approx72|t^{sp}|$ and $U_{xxxx}N_0=U_{yyyy}N_0\approx51|t^{sp}|$. Losses into the continum is the same as in Fig.~\ref{fig4}, i.e. $\Gamma_\nu/\hbar=0.1$ Hz ($\nu=0,\,x,\,y,\,xy,\,x0,\,0y$).} \label{fig5}
\end{center}
\end{figure}

For loading of $p$-band atoms we again use the pulsed acceleration (\ref{pulse}) but this time with $c=0.9999$. We see from Fig.~\ref{fig3} that the corresponding two dimensional Lorentzian is shifted away from $c=1$ more than the one dimensional Lorentzian. In the previous example we could keep $c=1$ and still attain a large population transfer. In two dimensions, on the other hand, utilizing $c=0.9999$ instead of $c=1$ increases the efficiency by several percent. Similarly, since the virtual processes are more important in this two dimensional situation we do not chose $T_r=\sqrt{2}t^{sp}$, but find $T_r$ numerically. The results for the full system evolution including the trap are shown in Fig.~\ref{fig5} as black curves. Losses $\Gamma_\nu$ are taken the same as for the one dimensional case. During the pulse switch off, the $p$-band population decreases with half a percent indicating non-adiabatic effects. 

\begin{figure}[ht]
\begin{center}
\includegraphics[width=8cm]{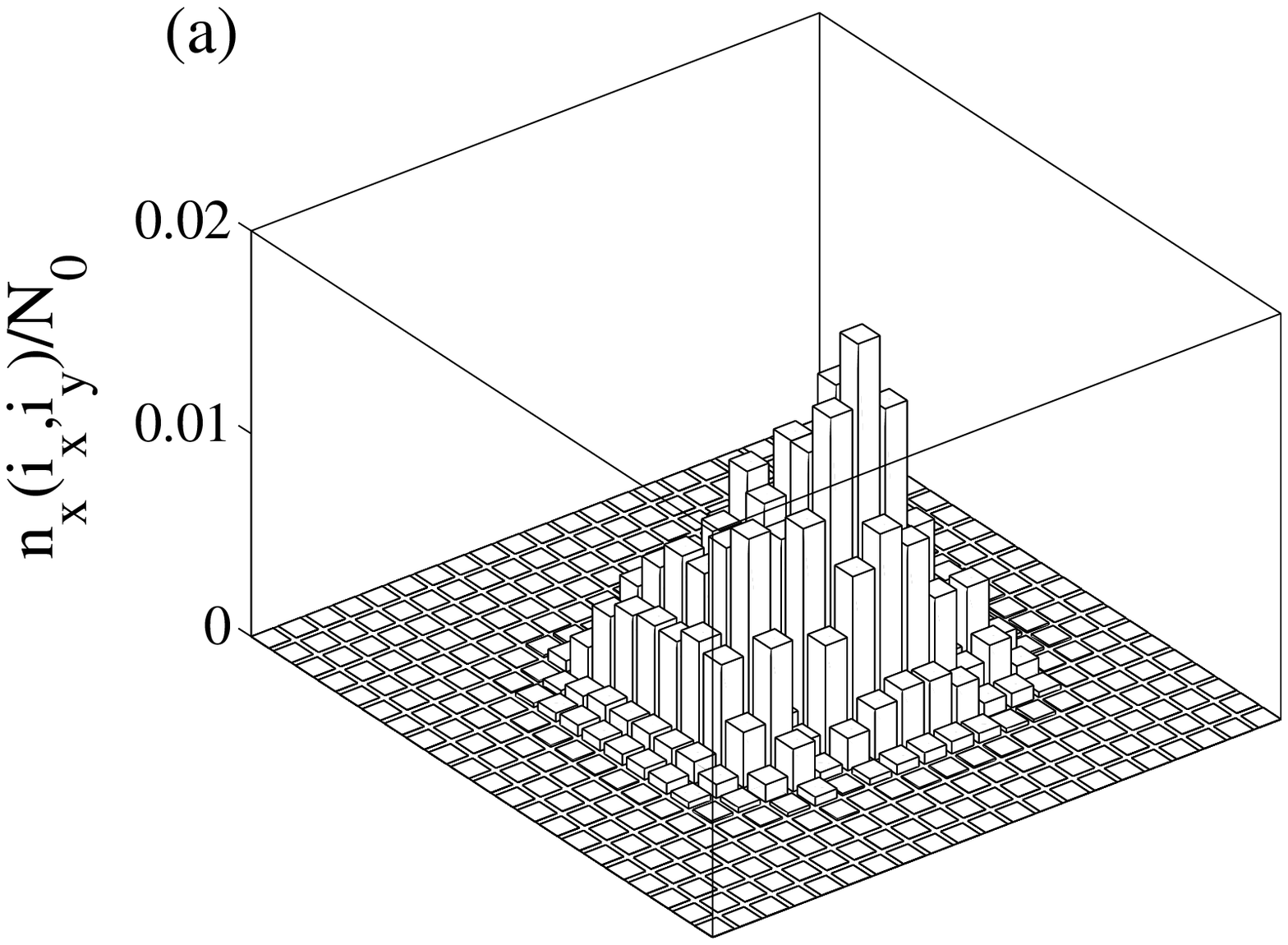}
\includegraphics[width=8cm]{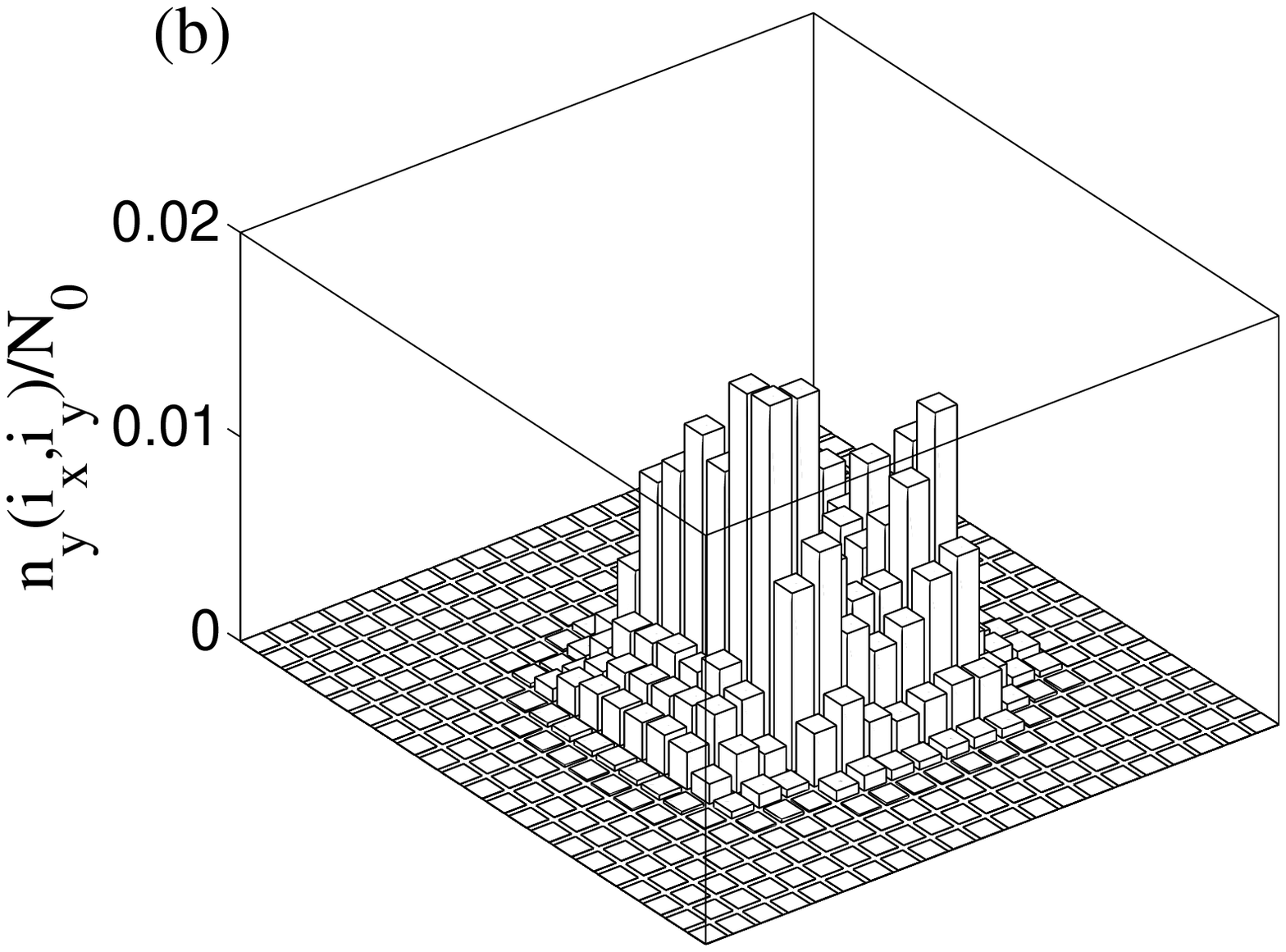}
\caption{Normalized atomic densities $n_x(i_x,i_y)/N_0$ (a) and $n_y(i_x,i_y)/N_0$ (b) on the $p$-band. As expected, the densities are mainly localized around the trap center.  } \label{fig6}
\end{center}
\end{figure}

Our numerical simulations allow us to extract other quantities, like the densities $n_x(i_x,i_y)$ and $n_y(i_x,i_y)$ which we display in Fig.~\ref{fig6}. Interestingly, the two densities are not identical and the $y$-density is slightly (5 $\%$) less populated. The origin of this asymmetry is not clear. Nevertheless, we note that since the two $p$-bands are degenerate and an imbalance of population between the two flavors does not change the energy of the system, i.e. the results are very sensitive to any small fluctuations. Atom-atom interaction might stabilize this, but as we are not focusing on the actual quantum state on the $p$-bands in this work, we do not examine this further.

In the interacting case, we do not solve the full system which would contain more than 1000 different interaction terms. Instead we use the fact that mainly the $s$- and $p$-bands are populated and therefore only include the interaction terms
\begin{equation}
\begin{array}{lllll}
U_{0000}, & & U_{xxxx}, & & U_{yyyy},
\end{array}
\end{equation}
corresponding to scattering processes that preserve the atomic orbital structure. Interactions that do not sustain orbital order, like $U_{00xx}$, $U_{00yy}$, and $U_{xxyy}$, are all smaller than the above. We find that moderate interaction can actually increase the population transfer. However, for extremely strong interaction, like in the one dimensional situation, a self-trapping effect sets in. In two dimensions, the system is less sensitive to interaction and therefore self-trapping becomes evident only for extremely strong atom-atom interactions. This is displayed in Fig.~\ref{fig5} as the pink curves. In this example, the strength of the interactions, $U_{\alpha\beta\gamma\delta}$, are one order of magnitude larger than the tunneling coefficient $t^{sp}$. The transition where interaction induced self-trapping becomes important is found to be relatively sharp. For example in Fig.~\ref{fig5}, for 20 $\%$ smaller interaction strengths the population transfer is increased to around 96 $\%$.

\begin{figure}[ht]
\begin{center}
\includegraphics[width=8cm]{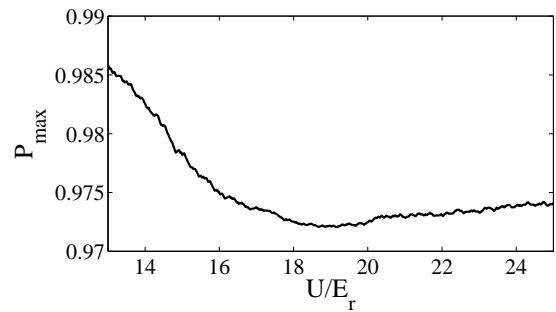}
\caption{The maximum population transfer $P_{max}$ from the $s$- to the $p$-bands as a function of the potential depth $V_L$. The interaction between the atoms is assumed zero, and the decay rate is taken the same for all $E_r$ and its value is as in Fig.~\ref{fig4}. } \label{fig7}
\end{center}
\end{figure}

We finish this section by investigating the transfer efficiency in terms of the potential depth. We consider non-interacting atoms and calculate the maximum population $P_{max}$ on the $p$-bands for various $V_L$. The results are presented in Fig.~\ref{fig7}. At smaller depths, we have noticed that the population transfer between the bands are no longer displaying a Rabi-like oscillation structure. Even at $V_L=10E_r$, mainly the $s$- and the $p$-bands are populated but their mutual dynamics is complex. In this irregular regime, at some occasions the $p$-band population can become very high while in the next moment decrease rapidly. At $V_L\sim18E_r$, the dynamics become more Rabi-like and from there on we encounter an increased loading efficiency with growing $V_L$'s. Experimentally one should pick a large $V_L$ in order to have a stable loading, but too large $V_L$ will imply long loading times causing heating and other incoherent processes to become important. 

\section{Conclusions}
We have shown that by properly accelerating optical lattices, bosonic atoms can be transfered from the lowest $s$-band to the first excited $p$-bands. The idea relies on creating a resonant coupling between the $s$- and the $p$-bands in the lattice. Since other transitions are off-resonant, the population mainly resides within these bands. With an efficiency of more than 95 $\%$ population transfer, the scheme seems as a good alternative compared to earlier methods like Raman pulses~\cite{raman}, ramped super-lattices~\cite{hemmerich1,hemmerich2}, or condensation into moving lattices~\cite{phillips}. 

The influence of atom-atom interactions was studied, and it was found that for strongly interacting gases self-trapping may deteriorate the $p$-band loading. This effect is considerably reduced in higher dimensions. Even in one dimension, the interaction must be very large in order to become influential. Experimentally, the interaction can be made weaker by the use of Feshbach resonances~\cite{feshbach} or opening up the trap, e.g. we consider a trapping frequency $\omega\sim250$ Hz and it could in principle be decreased by one order of magnitude~\cite{phillips}. Tunneling into the continuum was as well estimated to have a minor influence on the loading. In particular, for such deep lattices as $20E_r$ which we consider, the decay rate was found as 0.1 s$^{-1}$ which is an order of magnitude smaller than the characteristic loading rates. 

Throughout, we have tried to employ experimentally relevant parameters. For example, we considered $^{87}$Rb atoms and a lattice depth $V_L=20E_r$. This resulted in a loading-time of roughly 40 ms in two dimensions, which is within the coherence and heating times of a BEC~\cite{hemmerich1}. For the given parameters, the acceleration $a=F_0/m\approx500$ m/s$^2$ which is easily achieved experimentally~\cite{phillips}. The ability of the method for other lattice depths was as well discussed. Despite the fact that we consider experimentally relevant parameters and well developed techniques, there are possible sources that may lower the efficiency. For example, the method has a non-adiabatic character and it therefore relies on correctly adjusting the acceleration in terms of strength and duration. It follows from Fig.~\ref{fig3} that this can become a serious issue in one dimension where the population transfer is extra sensitive to the pulse strength. In two dimensions, this is less crucial, but one would probably need a control of the pulse strength within less than one percent. Another important aspect is the specific strengths of the tunneling coefficients obtained numerically from Wannier functions. The actual form of the Wannier functions, and especially their tails, greatly affects the values of the overlap integrals. Experimentally, the system might not occupy a true stationary state at $t=0$ and such corrections might affect the various $t_{\mathbf{ij}}^{\alpha\beta}$. It is also known that atom-atom interaction modifies the Wannier functions~\cite{wanbroad}, which will propagate into the tunneling coefficients. 

We considered a tight-binding approach and expanded the atomic field in terms of the three lowest band's Wannier functions. For the $s$-band, the tight-binding approximation is typically justified for $V_L>3E_r$~\cite{jonas}. For the higher excited bands, the validity of the tight-binding approximation is more stringent. This, however, is not assumed to cause any problems in our analysis since the higher bands are off-resonant beyond nearest neighboring sites, and even if they couple to these sites the effect is greatly suppressed by the large detuning. Therefore, by carefully controlling the experimental parameters, the present loading scheme seems to be experimentally feasible with current technology. Furthermore, we note especially two most recent experimental works~\cite{greiner} in which resonant tunneling between neighboring sites has been studied. Contrary to our work, Ref.~\cite{greiner} considers a strongly interacting system in the insulating regime, but it nevertheless demonstrate coherent population transfer between neighboring sites in tilted lattices.         

\begin{acknowledgments}

JL acknowledges support from the Swedish
government/Vetenskapsr{\aa}det.
\end{acknowledgments}


\begin{thebibliography}{999}
\bibitem{bands} The nomeclature has its origin in atomic physics. For more complicated lattice configurations, and thereby typically more involved band structures, it is not always appropriate to assign a label $s$, $p$, $d,\,...$ to each band.
\bibitem{com2} Naturally, when we consider one dimensional lattice, the lattice potential depth in the perpendicular $y$- and $z$-directions are assumed large such any dynamics in these directions can be froozen out. The same for two dimensions where $z$-motion is considered froozen.
\end{thebibliography}
\end{document}